\documentstyle[psfig]{mn}

\newif\ifAMStwofonts

\ifnfssone
  \newmathalphabet{\mathit}
  \addtoversion{normal}{\mathit}{cmr}{m}{it}
  \addtoversion{bold}{\mathit}{cmr}{bx}{it}
  \newmathalphabet{\mathbfit} 
  \addtoversion{normal}{\mathbfit}{cmr}{bx}{it}
  \addtoversion{bold}{\mathbfit}{cmr}{bx}{it}
  \newmathalphabet{\mathbfss} 
  \addtoversion{normal}{\mathbfss}{cmss}{bx}{n}
  \addtoversion{bold}{\mathbfss}{cmss}{bx}{n}
  \ifAMStwofonts
    \ifCUPmtlplainloaded \else
      \UseAMStwoboldmath
      \makeatletter
      \new@mathgroup\upmath@group
      \define@mathgroup\mv@normal\upmath@group{eur}{m}{n}
      \define@mathgroup\mv@bold\upmath@group{eur}{b}{n}
      \edef\UPM{\hexnumber\upmath@group}
      \new@mathgroup\amsa@group
      \define@mathgroup\mv@normal\amsa@group{msa}{m}{n}
      \define@mathgroup\mv@bold\amsa@group{msa}{m}{n}
      \edef\AMSa{\hexnumber\amsa@group}
      \makeatother
      \mathchardef\upi="0\UPM19
      \mathchardef\umu="0\UPM16
      \mathchardef\upartial="0\UPM40
      \mathchardef\leqslant="3\AMSa36
      \mathchardef\geqslant="3\AMSa3E
    \fi
  \fi
\fi 

\ifnfsstwo
  \DeclareMathAlphabet{\mathbfit}{OT1}{cmr}{bx}{it}
  \SetMathAlphabet\mathbfit{bold}{OT1}{cmr}{bx}{it}
  \DeclareMathAlphabet{\mathbfss}{OT1}{cmss}{bx}{n}
  \SetMathAlphabet\mathbfss{bold}{OT1}{cmss}{bx}{n}
  \ifAMStwofonts
    \ifCUPmtlplainloaded \else
      \DeclareSymbolFont{UPM}{U}{eur}{m}{n}
      \SetSymbolFont{UPM}{bold}{U}{eur}{b}{n}
      \DeclareSymbolFont{AMSa}{U}{msa}{m}{n}
      \DeclareMathSymbol{\upi}{0}{UPM}{"19}
      \DeclareMathSymbol{\umu}{0}{UPM}{"16}
      \DeclareMathSymbol{\upartial}{0}{UPM}{"40}
      \DeclareMathSymbol{\leqslant}{3}{AMSa}{"36}
      \DeclareMathSymbol{\geqslant}{3}{AMSa}{"3E}
    \fi
  \fi
\fi 

\ifCUPmtlplainloaded \else
  \ifAMStwofonts \else 
    \def\upi{\pi}
    \def\umu{\mu}
    \def\upartial{\partial}
  \fi
\fi

\title{Distribution of compact object mergers around galaxies}

\author[Bulik,Belczy{\'n}ski and  Zbijewski]{Tomasz Bulik$^1$,
Krzysztof Belczy{\'n}ski$^1$, and Wojciech
Zbijewski$^2$\\
Nicolaus Copernicus Astronomical Center, Bartycka 18,
00716 Warszawa,Poland\\
Department of Physics, Warsaw University, Ho{\.z}a 69, Warszawa,
Poland} 

\pagerange{\pageref{firstpage}--\pageref{lastpage}}
\pubyear{1999}

\begin{document}

\maketitle

\label{firstpage}

\begin{abstract}
 Compact object mergers are one of the
currently favored models for the origin of GRBs.  The discovery
of optical afterglows and identification of the nearest,
presumably host, galaxies allows the analysis of the distribution
of burst sites with respect to these galaxies.  Using a model of
stellar binary evolution we synthesize a population of compact
binary systems which merge within the Hubble time.  We include
the kicks in the supernovae explosions and calculate orbits of
these binaries in galactic gravitational potentials.  We present
the resulting distribution of merger sites and discuss the
results in the framework of the observed GRB afterglows. 

\end{abstract}


\section{Introduction}

The recent discoveries of X-ray afterglows of gamma-ray bursts
by the Beppo SAX satellite \cite{1997IAUC.6576....1C} have
revolutionized the approach to these phenomena.  For the first
time since their discovery 30 years ago \cite{Klebesadel68}
gamma ray bursts have been identified with sources at other
wavelengths.  In consequence optical afterglows have been
discovered \cite{1997IAUC.6584....1G}, which lead to
identification of host galaxies \cite{1997IAUC.6588....1G}.  At
the time of writing more than a dozen afterglows have been
identified.  The optical lightcurves of the GRB afteglows decay
as  a power law $\propto t^{-\alpha}$ with the typical values of
the index $\alpha = 1.1$ to $1.3$.  In some cases the host
galaxies have been found by observing the flattening of of the
lightcurve.  The underlying steady flux is identified as the
emission of the host galaxy.  In a few cases the host galaxies
themselves were found as extended objects.  This allowed to
measure the offset between the location of the gamma-ray burst
and the center of the host galaxy.  A list  five GRBs and the
offsets from their host galaxies  is shown in
Table~\ref{table1}.  The offsets are generally small and the
afterglows lie directly on the host galaxies.  In other cases
where the host galaxy have been only found from the flattening
of the lightcurve we know that the location of the host galaxy
and the GRB do not differ substantially from the simple fact
that the host galaxies have been identified.

There are two basic categories of theoretical models of the
central engines of gamma-ray bursts within the cosmological
model.  The first class connects gamma-ray bursts with mergers
of compact objects, e.g.  neutron stars and/or black holes. 
There exist numerous scenarios in this set of models, some of
them link GRBs with the coalescence of a black hole neutron star
binary \cite{1992ApJ...395L..83N}.  In other models like in the
recent paper by \cite{1998ApJ...505L.113K} the GRB events are
related to mergers of two  neutron stars.  Compact object merger
model provides enough  energy  to power a GRB, and it has been
showed in the numerical simulations \cite{1998ApJ...494L..53K}
that the coalescence may last up to a second.  The analytical
estimations of \cite{1998ApJ...503L..53P} extend this timescale
up to a minute. The second class of models
\cite{1998ApJ...494L..45P} relates GRBs to explosions of
supermassive stars, so called hypernovae. A direct prediction in
this class of models is that gamma-ray bursts are related  to
the star forming regions. 

The relation between the GRB location 
and the host galaxies does not have to be true in the models that
relate GRBs to compact object mergers. Tutukov and Yungelson
\shortcite{1993MNRAS.260..675T} have calculated the compact
object merger rates, and also found that compact object binaries
may travel the distances up to $1000\,$kpc before merging.  In a
more detailed study \cite{1998Bloom} calculated a population of
compact object binaries using the population synthesis method of
Pols and Marinus \shortcite{1994A&A...288..475P} and then
calculated the spatial distribution of mergers in the potentials
of galaxies for w few representative masses.  They found that approximately
$15$\% of mergers take place outside the host galaxies). They
have used the a Maxwellian kick velocity distribution with
$\sigma_v =190$~km~s$^{-1}$ \cite{1997MNRAS.291..569H}

In this work we use the the population synthesis code based on
\cite{Bethe1998} and extended  by Belczy{\'n}ski and Bulik
\shortcite{BB1998}.  We concentrate on the dependence of the
properties of the compact object binaries on the parameters used
in the population synthesis code.  We find that the most
important parameter that determines the population of compact
object binaries is the kick velocity a neutron star receives at
birth, however   this distribution is poorly known.  Iben and
Tutukov \shortcite{1996ApJ...456..738I} claim that the
properties of pulsars can be explained by only the recoil
velocities with no need for the kicks. Blaauw and Ramachandran
\shortcite{BlauwRama1998} find that a single kick velocity of
$200\,$km\,s$^{-1}$ suffices to reproduce the pulsar population.
Cordes and Chernoff \shortcite{1997ApJ...482..971C} proposed a
weighted sum two Gaussians:  80 percent with the width $175$\
km~s$^{-1}$ and 20 percent with the width $700$~km~s$^{-1}$.

We outline the model of the binary evolution and propagation in a
galactic potential in section~2.  The results of the calculation
are presented in section~3 and we discuss them in section~4.

\begin{figure*} 

\psfig{width=\textwidth,file=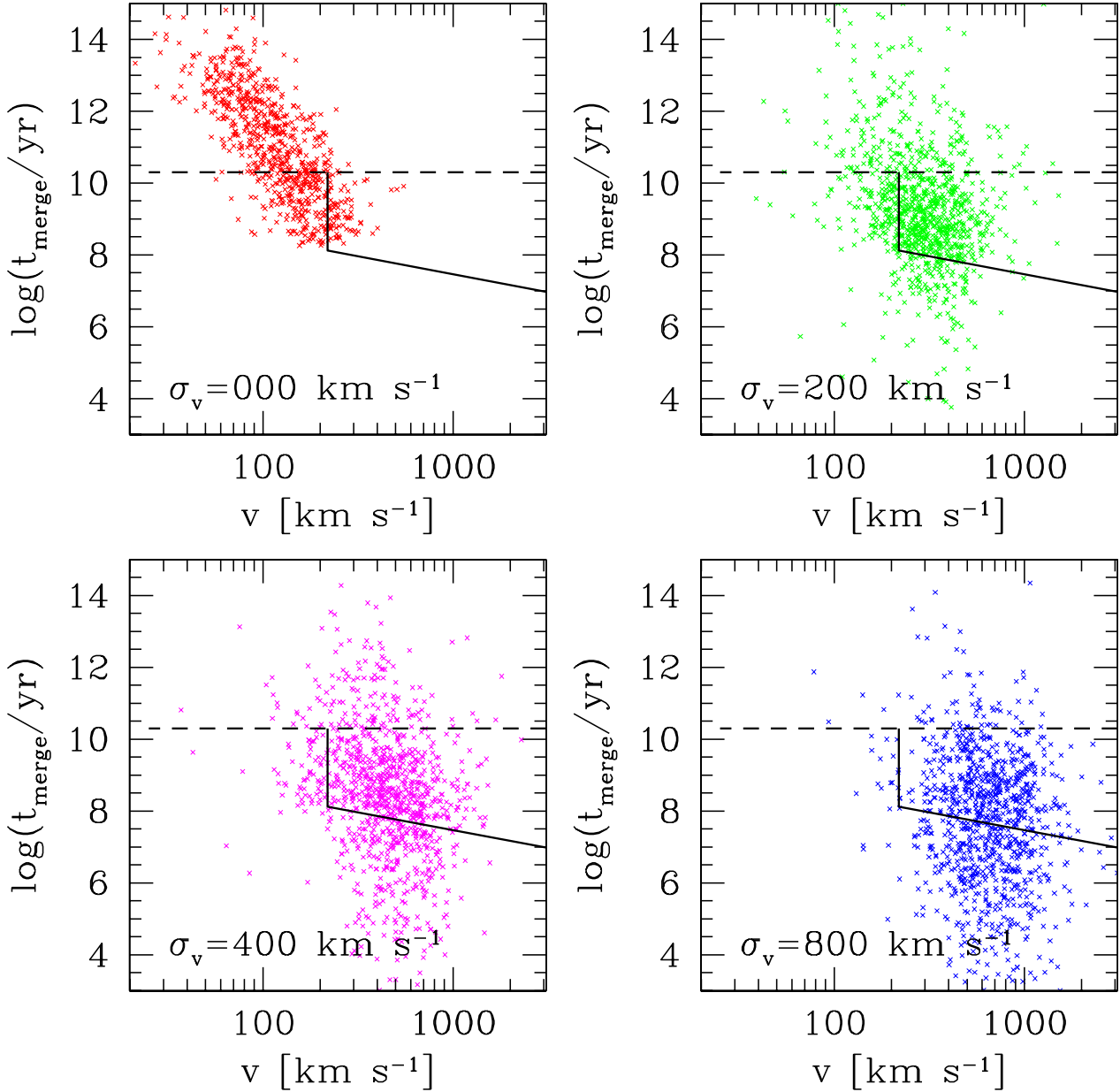}
\caption{Distribution of compact object system in the velocity
versus lifetime (time to merge).  The top left panel shows the
case when there is no kick velocity, the top right panel shows
the case $\sigma_ = 200\,$km~s$^{-1}$, the bottom left panel
$\sigma_ = 400\,$km~s$^{-1}$, and the bottom right panel is for
$\sigma_ = 800\,$km~s$^{-1}$. The horizontal dashed line
corresponds to the Hubble time (15Myrs). In the region for 
$t_{merge} < 15\,$M we present two solid lines: the vertical
corresponding to $v=200\,$km~s$^{-1}$ - approximately the escape
velocity from a galaxy, and the line corrsponding to a constabt
value of
$v\times t_{merge} =30\,$kpc. Together these lines define the
region in the parameter space with systems 
that can escape from the host galaxy. }
  \label{vt} 
\end{figure*}

\section{Model}

\subsection{Binary evolution}

In order to study the spatial distribution of compact object
mergers we use the population synthesis method.  We use the
population synthesis code \cite{BB1998} which concentrates on
the population of massive star binaries, i.e.  those that may
eventually lead to formation of compact objects and compact
object binaries.  We include the evolution of the binaries due to
interaction and mass transfer and also the kicks that a newly
born neutron star receives in supernova explosion.  A binary may
be disrupted in each of the supernova events.  The surviving
binaries obtain center of mass velocities, which change their
trajectories and may even eject them from their galaxy.

While the evolution of single stars depends only on their mass
and metallicity the evolution of binaries is also a function the
initial orbit (semimajor axis $a$, and eccentricity $e$) of the
two stars.  We assume that the distribution of the initial
parameters can be expressed as a product of distributions of four
parameters:  the larger star (primary) mass $M$, the mass ratio
of the less massive to the more massive star in the binary
$q$, and the orbital parameters $a$ and $e$, i.e that this
quantities are independent.  The distribution of primary masses
used here is \cite{Bethe1998} \[ \Psi(M) \propto M^{-3/2}\, , \]
and we adopt a flat distribution of the mass ratio $q$.  The semi
major axis distribution is scale invariant, i.e.  \[ \Gamma(a)
\propto a^{-1} \] with the limits $6R_\odot < a < 6000R_\odot$,
and we draw the eccentricity from a distribution $\Xi(e) = 2e$.

We assume that the kick velocity distribution is a three
dimensional Gaussian, 
and parameterize it with its width
$\sigma_v$, i.e.
 \begin{equation}
 p(v)\,
 = {4\over\sqrt\pi} \sigma_v^{-3} 
 v^2 \exp\left( -{v^2 \over \sigma_v^2}\right)\, .
 \end{equation}
 We generate population of compact object binaries
for a few values of $\sigma_v$ in order to asses the sensitivity
of our results to this parameter.

\begin{table}
\caption{GRBs with measured offsets from the centers of their host galaxies.}
\begin{tabular}{l|l|r|l}
 GRB    &  redshift $z$   &  Offset $\Delta \Theta $ & Reference \\ \hline
970228  &   ???  &  $ 0.30$'' & \cite{1997ApJ...489L.127S}       \\
970508  &  0.835 &  $ 0.01$'' & \cite{1998Fruchter}              \\
971214  &  3.42  &  $ 0.06$'' & \cite{1998Natur.393...35K}       \\
980703  &  0.966 &  $ 0.21$'' & \cite{1998ApJ...508L..21B}      \\
990123  &  1.60  &  $ 0.60$'' & \cite{GCN256}
 \end{tabular}
 
\label{table1}
\end{table}

We describe the mass transfer in the common envelope evolution by
the common envelope parameter $\alpha_{CE}$ (see e.g.
\cite{1991A&A...249..411V}), and we use an intermediate value of
$0.8$ for this parameter.  In this type of evolution the more
massive star looses its envelope and becomes a helium star with
mass approximately 30\% of its initial value.  The $\beta$
parameter which describes the specific angular momentum of the
material expelled from the binary in the Roche lobe overflow
phase is set to $\beta=6$ \cite{1994A&A...288..475P}.  Accretion
onto a neutron star in a binary is treated as Bondi-Hoyle
accretion and we use the formalism developed by \cite{Bethe1998}
to find the amount of mass accreted onto the neutron star, and
the final orbital separation.  Systems with nearly equal masses
evolve at the similar speed, and loose the common envelope,
shrinking their orbit at the same time.  For a more detailed
description of the population synthesis code see \cite{BB1998}.

We assume that a neutron star with mass of $1.4\, M_\odot$ is
formed in each supernova explosion.  We draw a random time in the
orbital motion to obtain the position on the orbit when the
supernova explodes.  The remaining mass of the envelope is ejected
from the system, and the newly formed neutron star receives a
kick.We verify whether the system is still bound after the
explosion.  For bound systems we find the parameters of the new
orbit and the kick velocity the whole binary receives 
\[ \Delta
\vec V = {M_2^i - M_2^f \over M_1 + M_2^f }(\vec v_2^i + \vec
v_{kick} ) \] 
where $M_1$ is the mass of the companion, $M_2^i,\,
M_2^f$ are the initial and final masses of the supernova, $\vec
v_2$ is the orbital velocity of the supernova at the time of
explosion. After each supernova expolsion we verify whther the
system survives as a binary.

A compact object binary loses its energy through gravitational
radiation.  The time to merge is \cite{Peters1964}
\begin{equation} 
t_{mrg}= {5c^5a^4 (1-e^2)^{7/2} \over 256G^3 Mm
(M+m)} \left( 1+ {73\over 24} e^2 + {37\over 96} e^4\right)^{-1}\
, 
\end{equation} where $a$ is the semi major axis of the orbit,
$e$ is its eccentricity, and $M,\, m$ are the masses of the
compact objects.

\begin{figure}
\psfig{width=0.45\textwidth,file=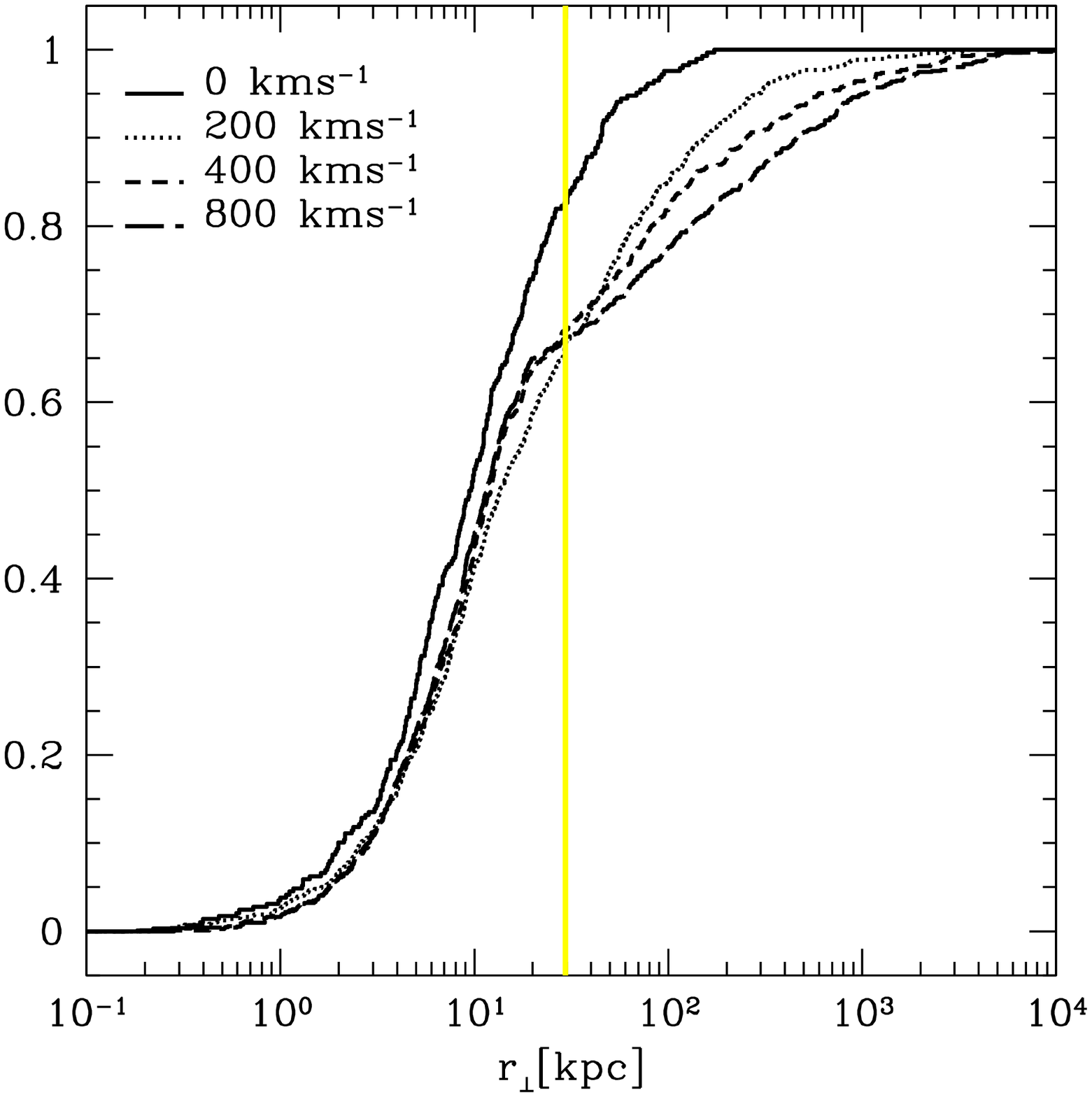}
\caption{The cumulative distributions of the projected distances
$r_\perp$ of compact object binaries in a potential of a large galaxy.
The curves correspond to the four values of the kick 
velocity$\sigma_ = 0,\,200,\,400,\,800\,$km~s$^{-1}$. The
vertical line corrsponds to the distance of $30\,$kpc from the
center of galaxy.}
\label{distuz}
\end{figure}

\begin{figure}
\psfig{width=.45\textwidth,file=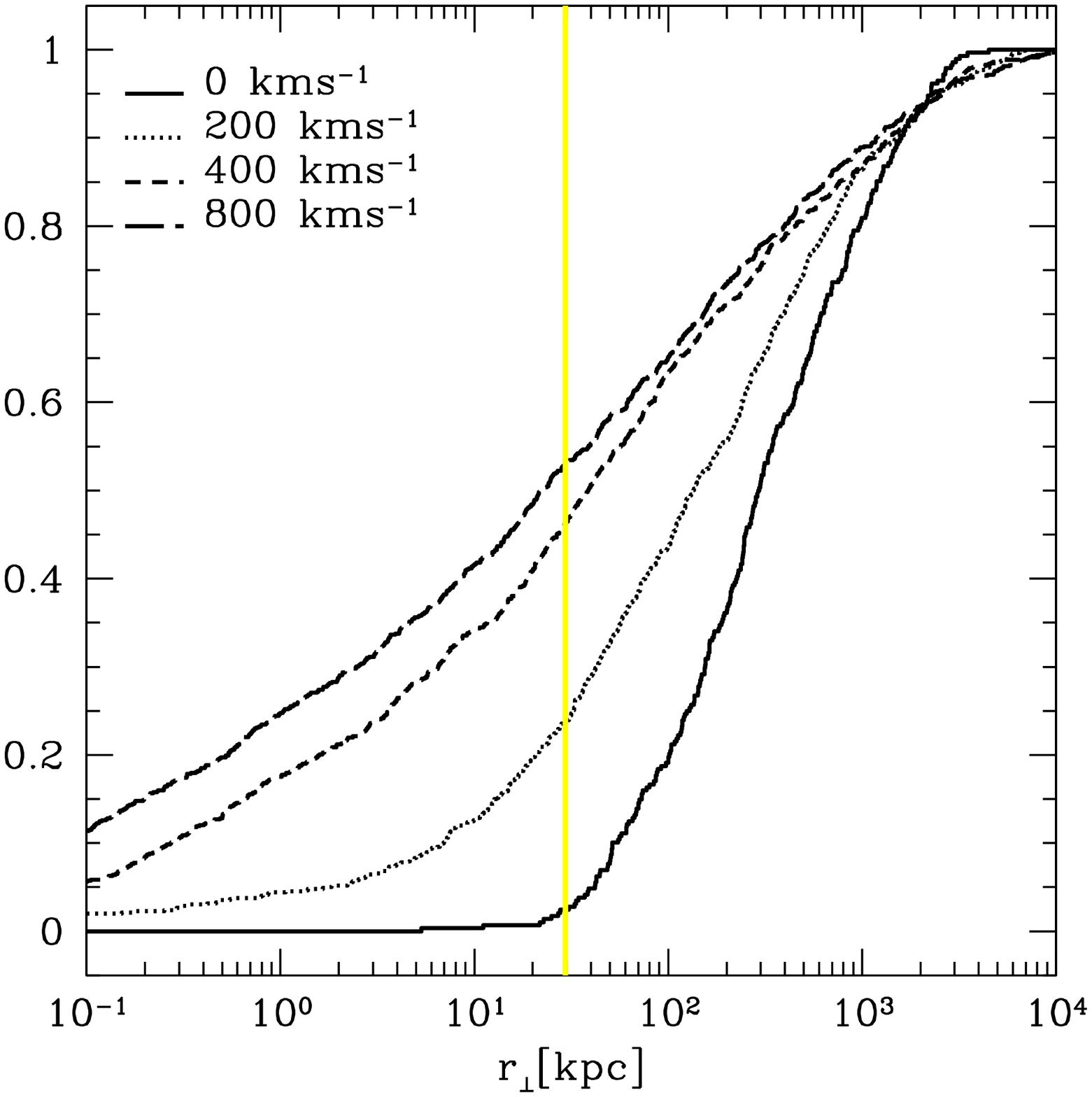}
\caption{The cumulative distributions of the projected distances
$r_\perp$ of compact object binaries in empty space.
The curves correspond to the four values of the kick 
velocity$\sigma_ = 0,\,200,\,400,\,800\,$km~s$^{-1}$.The
vertical line corrsponds to the distance of $30\,$kpc from the
point where the systems originate.}
\label{distvt}
\end{figure}

\subsection{Orbit in  Potentials of Galaxies}

Since little is known about the host galaxies of gamma-ray bursts,
in particular of their types and masses, we will present two
extreme cases:  (i) propagation in a potential of large spiral
galaxy like the Milky Way, and (ii) propagation in empty space,
corresponding to GRBs originating e.g. in globular clusters.
In the latter case we assume that all binaries originate in one
point, and travel due to kicks described above and there is no
gravitational potential.

The potential of a spiral galaxy can be described as a sum of
three components:  bulge, disk, and dark matter halo.  A
convenient way to describe the Galactic potential has been
proposed by \cite{1975PASJ...27..533M}, while a series of more
detailed models were constructed by \cite{1989MNRAS.239..571K}
and used in modeling the galactic halo population of neutron
stars \cite{1995Ap&SS.231..373B,1998ApJ...505..666B}.  The
\cite{1975PASJ...27..533M} potential for a galactic disk and
bulge is \[ \Phi(R,z) = {GM\over \sqrt{ R^2 + (a_i + \sqrt{ z^2
+ b_i^2})^2}} \] where $a_i$ and $b_i$ are the parameters, $M$
is the mass, and $R=\sqrt{x^2 = y^2}$.  The dark matter  halo
potential is spherically symmetric \[ \Phi(r) = - {G M_h\over
r_c} \left[ {1\over 2} \ln\left( 1 + {r^2\over r_c^2}\right) +
{r_c\over r} {\mathrm{atan}} \left( r\over r_c\right) \right] \]
corresponds to a mass distribution $\rho = \rho_c/[1 +
(r/r_c)^2]$.  The mass of such halo is infinite, so we introduce
a cutoff value $r_{cut} = 100\,$kpc above which the density of
the halo falls to zero. While the details of the model of
galactic potential are not important for this study we have to
adopt a particular value of the masses and sizes of each of the
components.  We use the values of the parameters as determined
by \cite{1991ApJ...381..210B} for the Milky Way:  $a_1 =0\,$kpc,
$b_1 = 0.277\,$kpc, $a_2 = 4.2\,$kpc, $b_2 = 0.198\,$kpc, $M_1 =
1.12\times 10 ^{10}\,M_\odot$, $M_2 = 8.78\times
10^{10}\,M_\odot$, $r_c = 6.0\,$kpc, and $M_h = 5.0 \times
10^{10}\,M_\odot$.

We assume that the distribution of binaries in our model galaxy
follows the mass distribution in the young disk
\cite{1990ApJ...348..485P}, that is \[ P(R,z)\,dR\, dz =
P(R)dR\, p(z)d(z)\, .  \] The radial distribution is exponential
with \[ P(R) \propto e^{-R/R_{exp}} R\, \] with $R_{exp} =
4.5\,$kpc and extends to $R_{max}= 20\,$kpc.  The vertical
distribution is $p(z) \propto e^{-z/z_{exp}}$ and $z_{exp} =
75\,$pc.  We note that this is not a self consistent approach:
the density inferred from the disk potential is not the same as
the density of binaries.  However in this work we are not
interested in determining high accuracy positions around the host
galaxy, and rather with an estimate of the general properties of
the distribution of compact object mergers.

Each binary moves initially with the local rotational velocity in
the galactic disk.  After each supernova explosion we add an
appropriate velocity, provided that the system survives the
explosion. We calculate the orbit of each system until it merger
time provided that the merger time is smaller than the Hubble
time (15 Gyrs here).

\begin{figure*}

\begin{tabular}{lr}
\psfig{width=0.45\textwidth,file=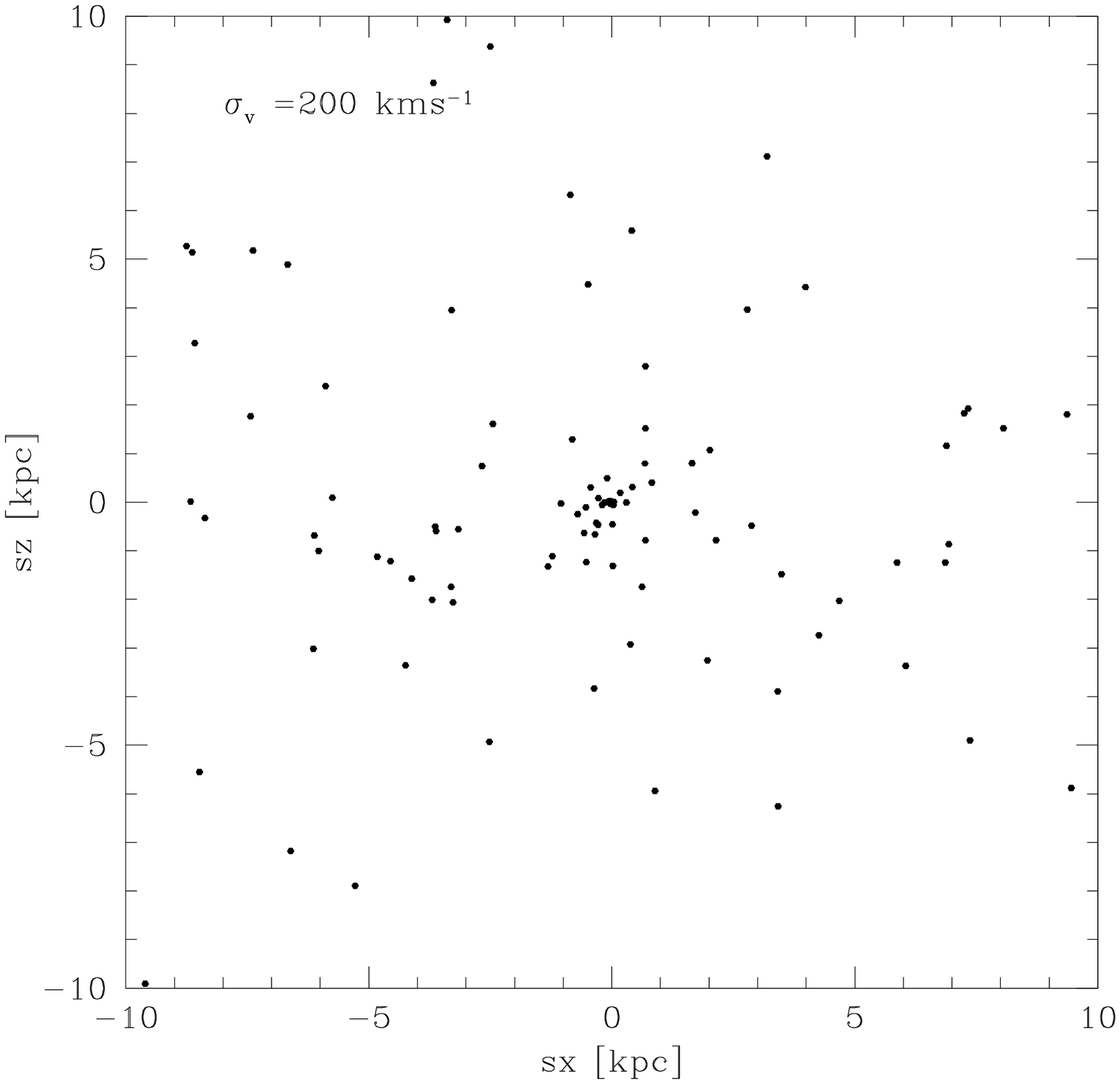}
&
\psfig{width=0.45\textwidth,file=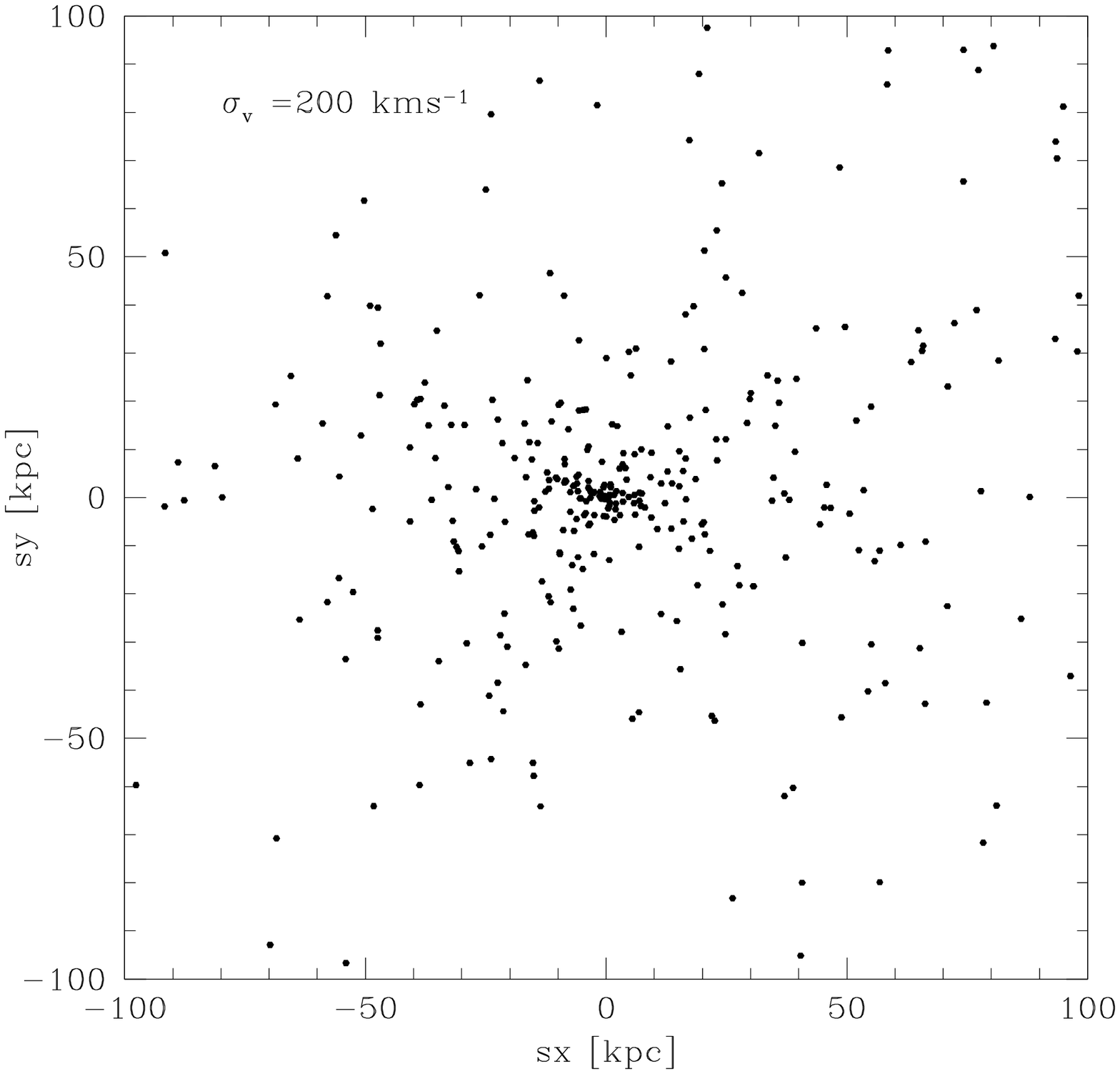}
\\
\psfig{width=0.45\textwidth,file=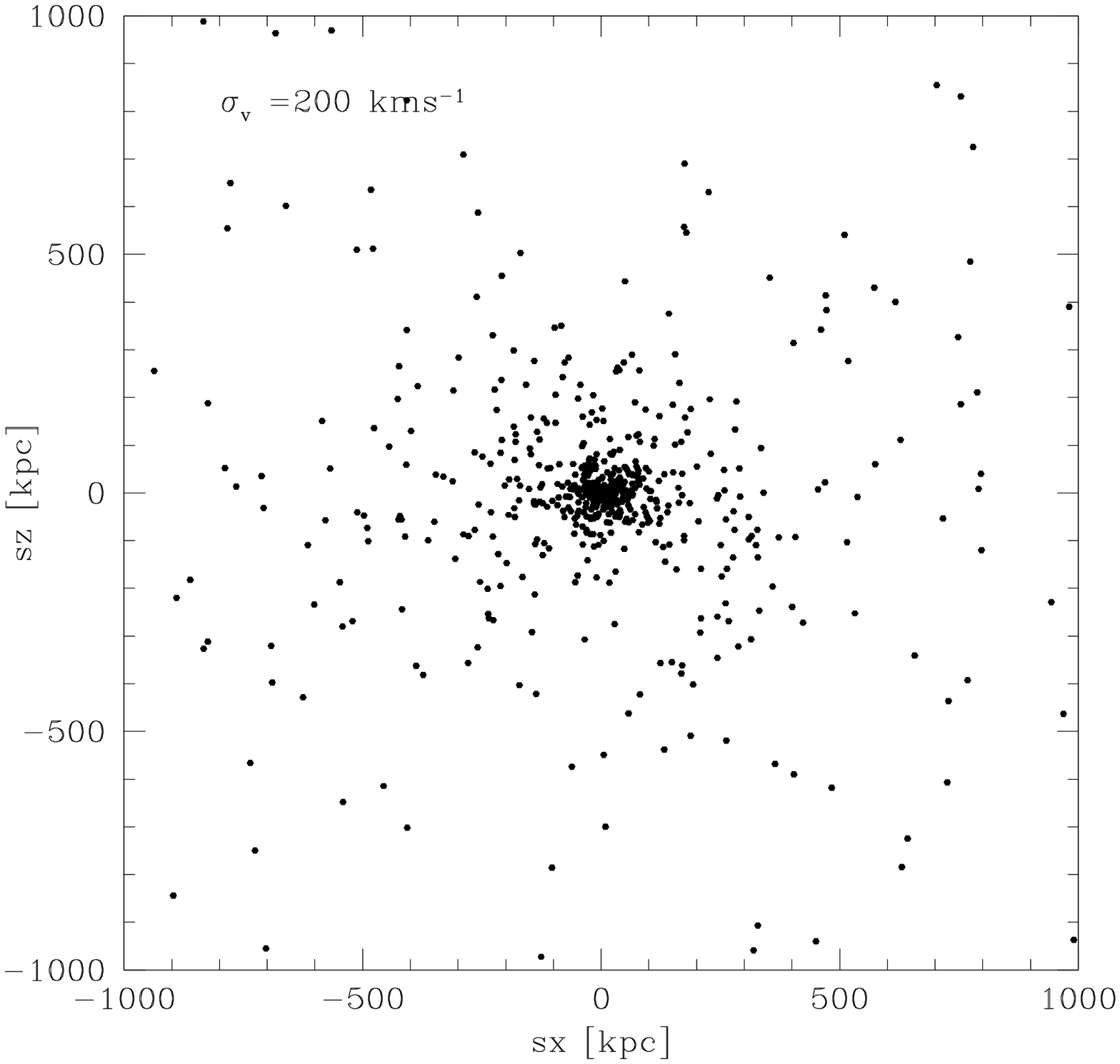}
&
\psfig{width=0.45\textwidth,file=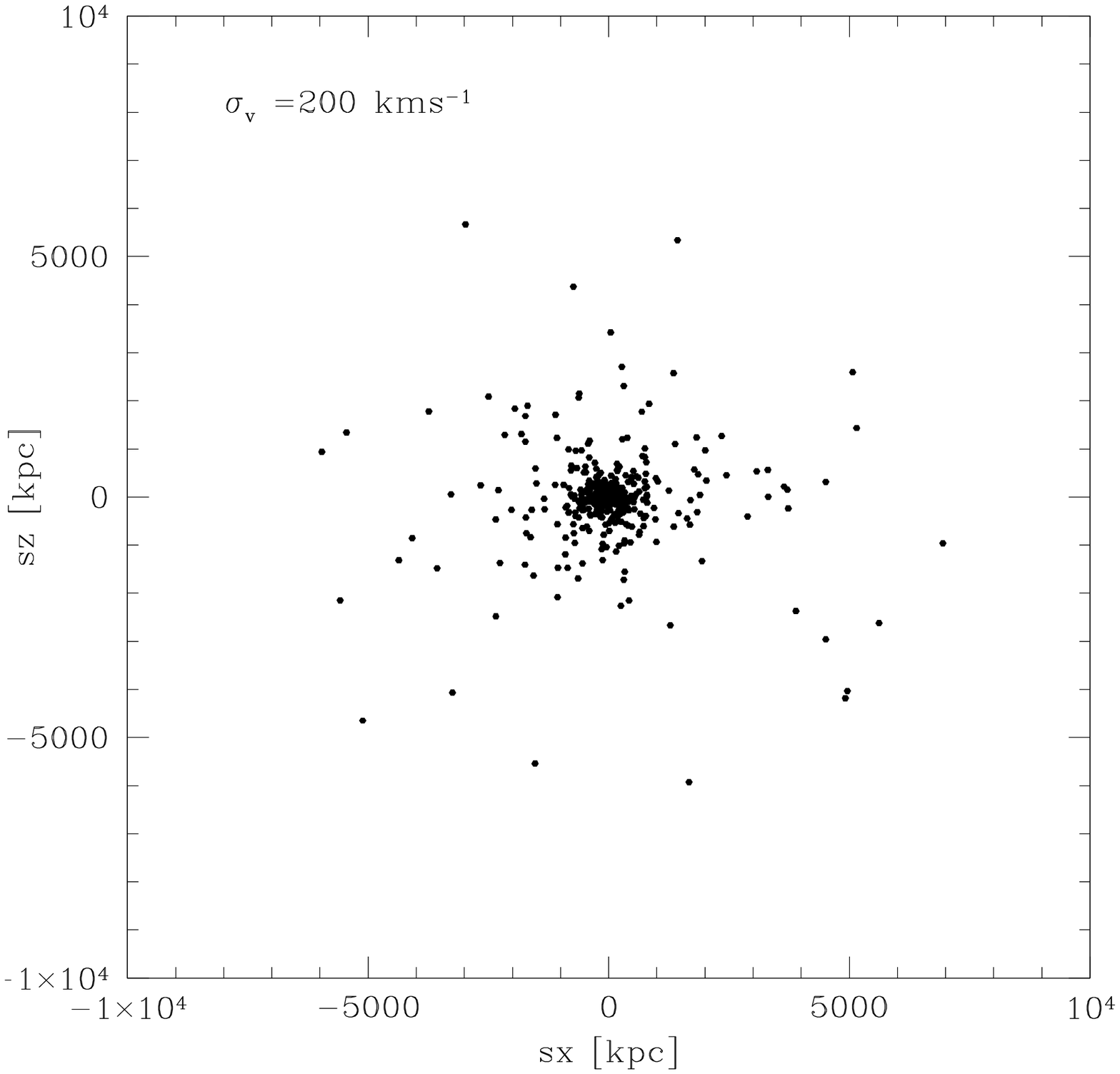}
\end{tabular}
\caption{The distribution of compact object mergers 
around a small galaxy (approximated by empty space).
The region shown in the plots increases from 
$10\,$kpc in the  top left panel,
through $100\,$kpc in the top right panel, to
$1000\,$kpc in the bottom left panel and to finally
to $10\,$Mpc in the bottom right panel.}
\label{empty-dot}
\end{figure*}

\begin{figure*}

\begin{tabular}{lr}
\psfig{width=0.45\textwidth,file=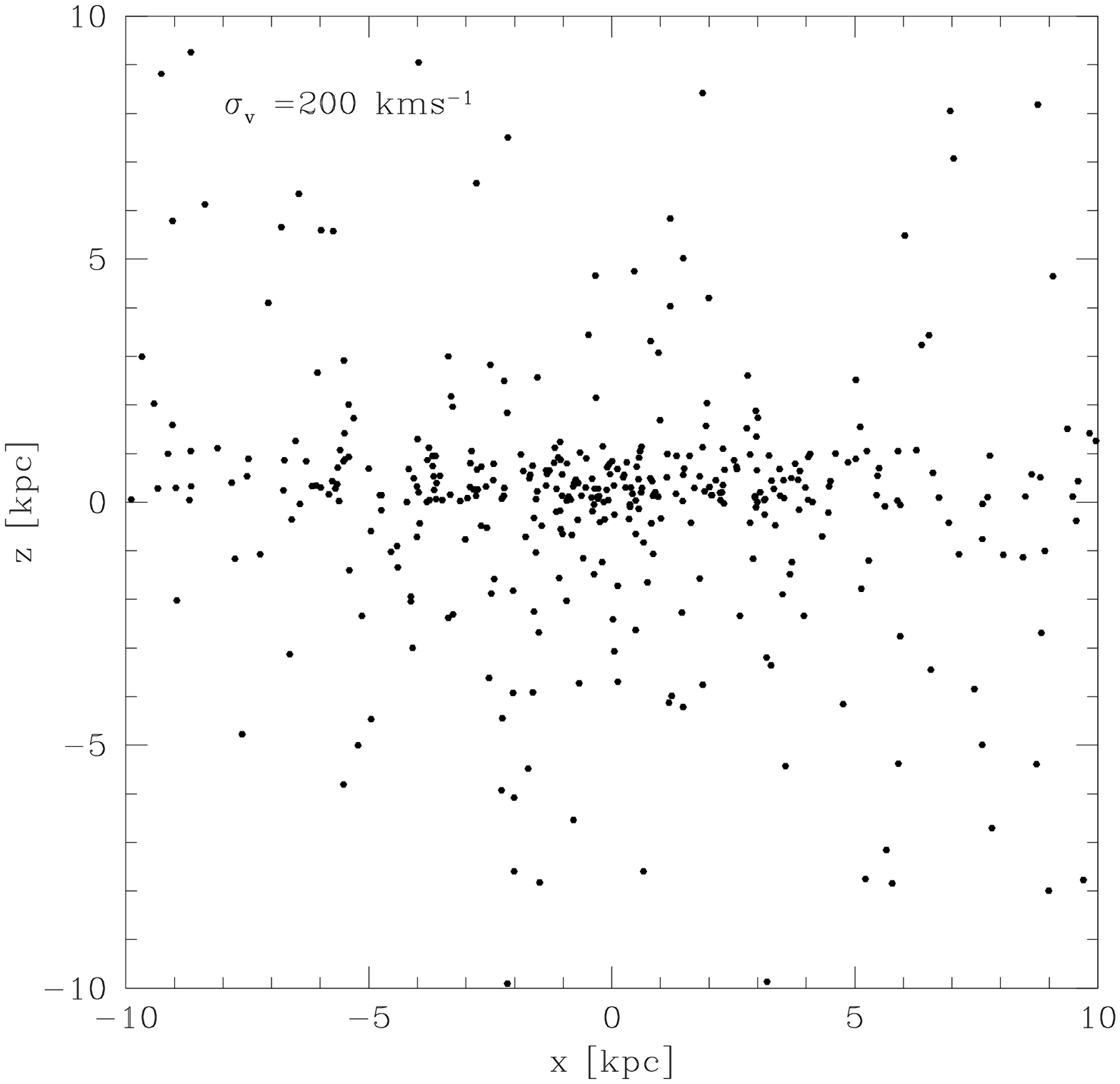}
&
\psfig{width=0.45\textwidth,file=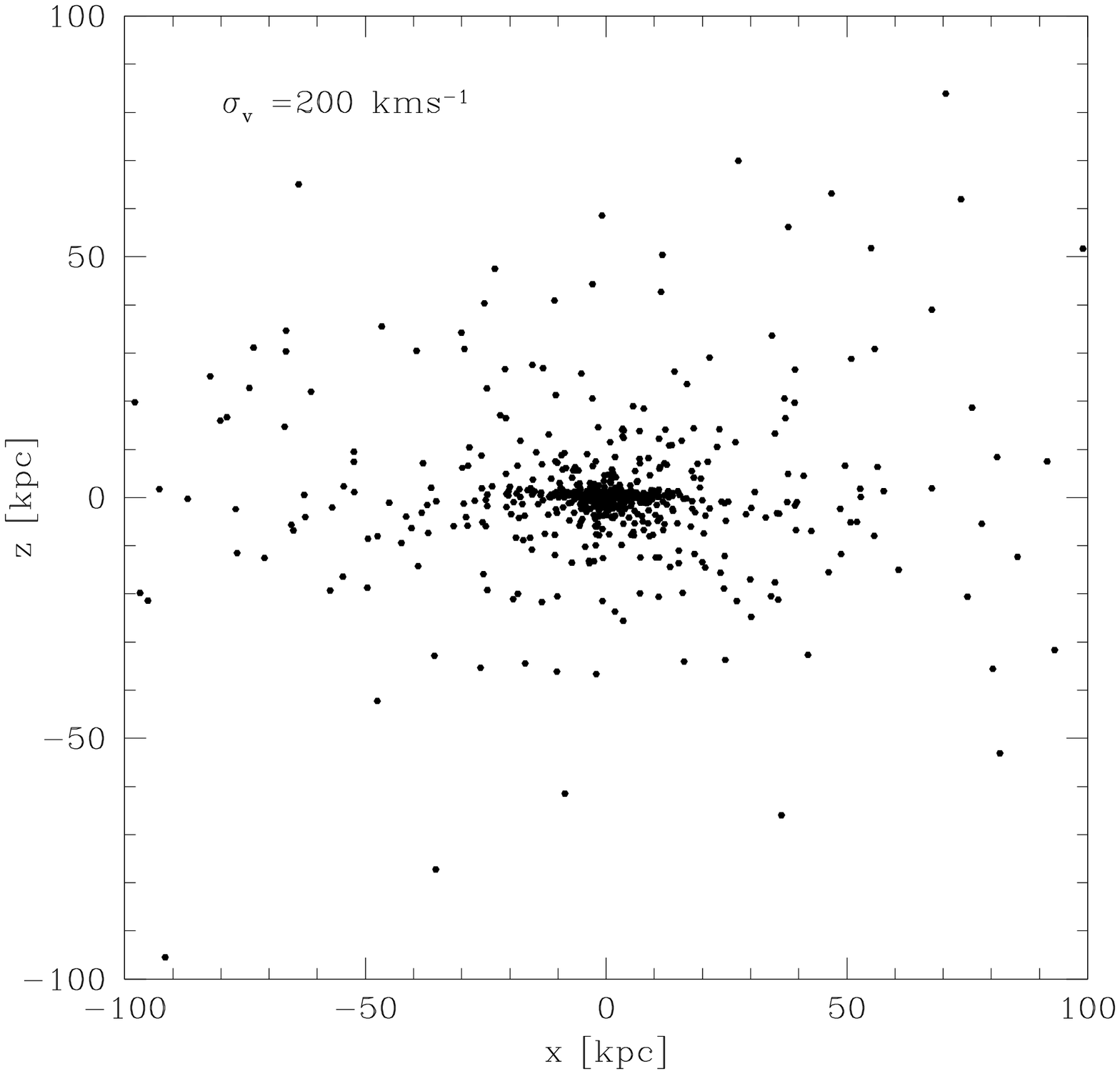}
\\
\psfig{width=0.45\textwidth,file=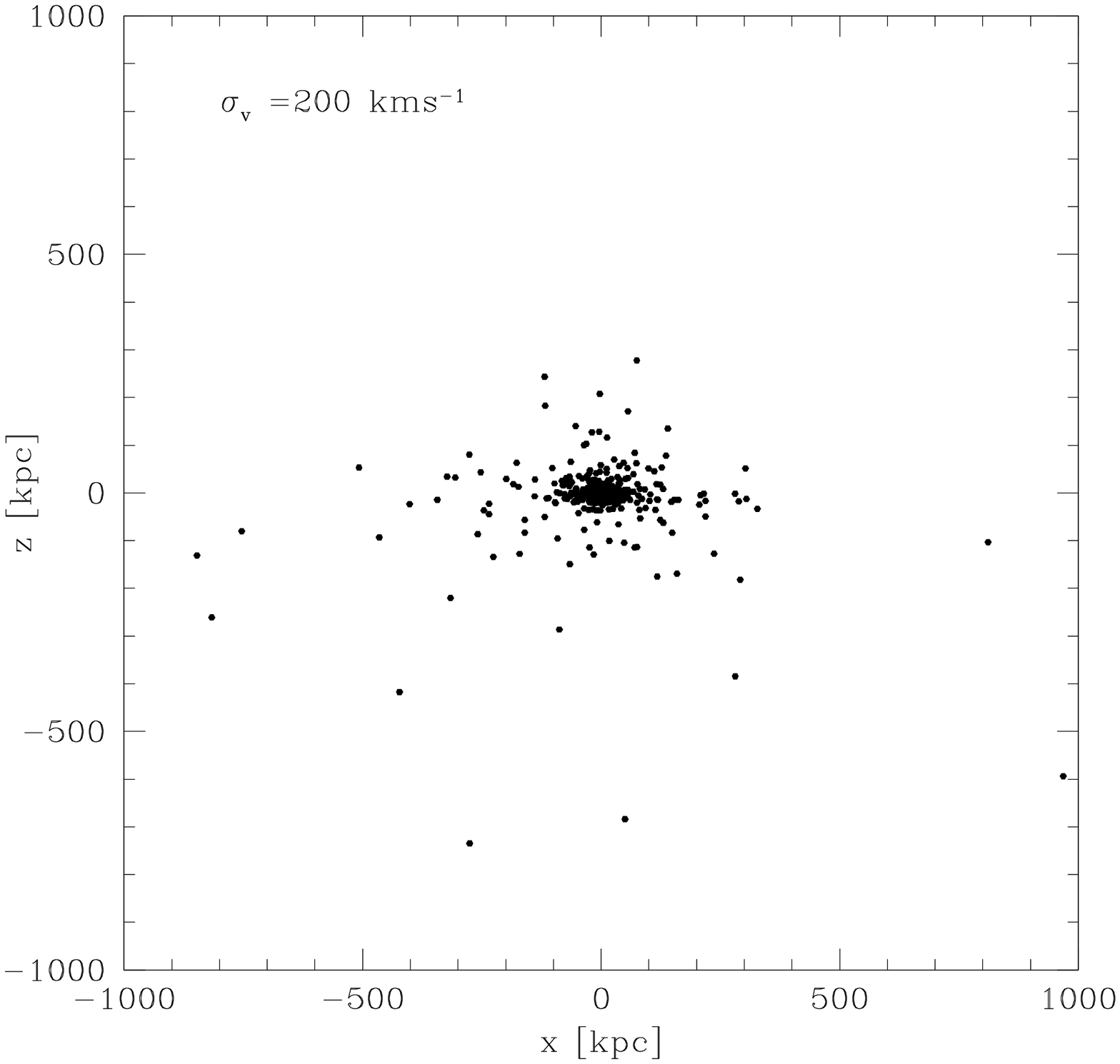}
&
\psfig{width=0.45\textwidth,file=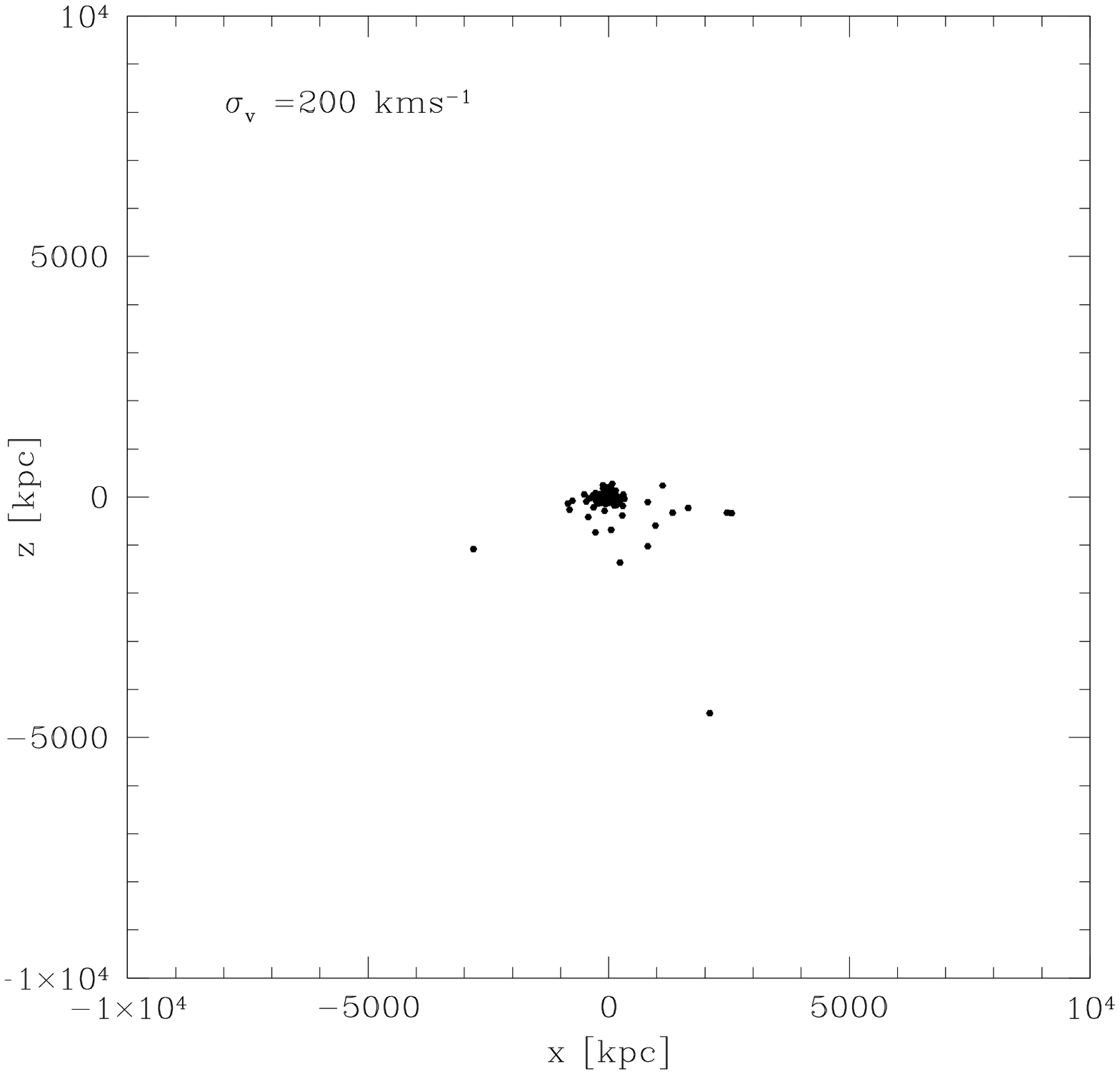}
\end{tabular}
\caption{The distribution of compact object mergers 
around a massive galaxy.
The region shown in the plots increases from 
$10\,$kpc in the  top left panel,
through $100\,$kpc in the top right panel, to
$1000\,$kpc in the bottom left panel and to finally
to $10\,$Mpc in the bottom right panel.}
\label{gala-dot}
\end{figure*}

\section{Results}

The kick velocity distribution is not very well known.
Therefore, we use the population synthesis code with four values
of the kick velocity distribution width:  with no kick velocities
$\sigma_v =0\,$km~s$^{-1}$, and with $\sigma_v = 200,\, 400,\,
800\,$km~s$^{-1}$.  This covers the range of values this
distribution is likely to have.  This the same approach as
adopted in our previous work \cite{BB1998}.

The binaries receive kicks for two reasons.  First, the envelope
of the supernova is lost from the system and it 
carries away some momentum.
Thus even in the case when there is no kick velocity a binary
achieves an additional velocity \cite{1960BAN....15..265B}.
Second, if the supernova explosion is asymmetric both the newly
formed compact object may receive a kick velocity which affects
the orbit of the binary after the explosion as well as its center
of mass velocity.  The fate of a binary system in a supernova
explosion depends on the value and direction of the kick
velocity, on the orbital phase at which the explosion occurs, and
on the parameters of the binary: the masses and orbital
parameters $a$, and $e$.

We present the population of compact object binaries in the plane
spanned by the center of mass velocity ofter the second
supernova expolsion  and time to merge in Figure~\ref{vt}.
The orbital \cite{1960BAN....15..265B} effects are isolated and
shown in the top left panel of Figure~\ref{vt}, where we present
the results of the simulation with $\sigma_v=0$.  There is a tail
of long lived systems with lifetimes much longer than the Hubble
time and small velocities, stretching outside of the boundaries
of the plot to the lifetimes even of $10^{20}\,$years.  These
systems originally had large orbital separations, and hardly
interacted in the course of their binary lifetime.  
In the case when there are no kicks the center of mass velocity
of the comapct object binary depends on the amount of mass lost
in the supernova explosion. In the extreme case of large mass
loss, the center of mass velocity approaches the orbital velocity
at the moment of supernova explosion, and it can never exceed it.
The velocity of the system increases with increasing mass loss,
however the systems that loose too much mass become unbound.
This is why the lower part of the plot below $t_{merge} \approx
10^8\,$years is empty.  With increasing the kick velocity also
the typical velocity of a system increases and there appear short
lived systems in tight orbits.  They can now survive a large mass
loss when the kick velocity has a favorable direction.  Thus as
the kick velocity is increased only the tightly bound systems
(with short merger time) can
survive the supernova explosions.  Another effect of the kick
velocity is that the long lived systems with $t_{merge}$ much
longer than the Hubble time, which were present in the case
$\sigma_v=0\,$km~s$^{-1}$ disappear.  The typical velocity of a
system increases with the kick velocity.  However, the population
of the comapct merger binaries is not much affected when the kick
velocity becomes large, e.g.  changing the kick velocity
distribution width from $\sigma_v=0\,$km~s$^{-1}$ to
$200\,$km~s$^{-1}$ produces a much stronger effect than going
from $\sigma_v=400\,$km~s$^{-1}$ to $800\,$km~s$^{-1}$.  Most of
the systems are disrupted by such high velocities, and the
surviving ones are only those for which the kick are not so
large and have a favorable direction.

Another effect of increasing the kick velocity is that the
typical lifetime of a system becomes smaller.  When the kick
velocity is large only very tight, and/or highly eccentric
systems survive, hence the typical lifetime of compact object
binaries decreases.  It should be noted the typical center of
mass velocity of the compact object binaries increases roughly
linearly with the kick velocity, while the lifetime decreases
approximately exponentially.

In Figure~\ref{vt} we also plot following \cite{1998Bloom} the
lines corresponding to the Hubble time (the dashed line), and we
mark the region with the stars that will escape from  a galactic
potential. In order to escape a binary must satisfy the
following conditions: (i) it has to have a velocity larger than
the escape velocity, (ii) the distance $vt_{merge}$must be
larger than the size of the galzxy. We also draw the line  at
$t_{mrg}=15\,$Myrs, to denote the systems that merger within the
Hubble time. All the systems to the right of the solid line in
Figure~ref{vt} have velocities  above $200\,$km~s$^{-1}$, and
live long enough to travel further than $30\,$kpc. We should
also note that although each panel in Figure~\ref{vt} contans
$10^3$ systems, the production rate of compact object binaries
decreases exponentially with the increasing kick velocity
(see eq. 13 in Belczy{\'n}ski and Bulik \shortcite{BB1998}).

In Figures~\ref{distuz} and~\ref{distvt} we present the
cumulative distributions of the projected distance from the
center of the host galaxy in case (i) and from their the place of
birth  in case (ii), respectively, of the systems that merge
within the Hubble time. When the binaries propagate in the
potential of a large galaxy the kick velocity only weakly 
influences the  the distribution of the mergers. Below the
radius of  $10\,$kpc the distribution is determined by the
potential well. This is where all short lived and slow systems
merge.  There exists however a tail of high velocity, long
lived  systems (see Figure~\ref{vt}) that manage to escape. The 
escaping fraction is a weak function of the kick velocity.
Typically the number of systems that merge further than
$30\,$kpc from the center of the host galaxy is $30$\%, except
for the unphysical case of no kick velocities when it drops
below $20$\%.

In the other extreme case  of small host galaxies for which we 
neglect the gravitational potential the escaping fraction can be
even larger. The escaping fraction decreases from $80$\% for the
kick velocity $\sigma_v = 200\,$km~s$^{-1}$ to about $50$\% for
$\sigma_v = 800\,$km~s$^{-1}$. The reason for such behavior is
clearly seen from Figure~\ref{vt}. In the product  of center of
mass velocity and time to merge the dominant role is played by
the fast decrease of the time to merge with the increasing kick
velocity $\sigma_v$.

These quantitative results are visualized in
Figures~\ref{empty-dot} and~\ref{gala-dot}. Here we show the
distribution  of $10^3$ mergers around massive galaxy and in the
empty space. We are showing four panels 
that cover the scales from $10\,$kpc to $10\,$Mpc. 
In the case of the propagation in a massive galaxy we are showing
the projection in the plane of the galactic disk so the
effects of the rotational velocity and the asymmetry of the
potential well are visible. Both calculations have been done for
the case of the kick velocity distribution width 
$\sigma_v = 200\,$km~s$^{-1}$.

\section{Conclusions}

We find that a significant fraction  i.e. more than  $20$
percent  of compact object mergers take place outside of the
host galaxies. The figures obtained for our case of no
gravitational potential should be considered  as an upper limit
only. In contrast the observations show that the GRB afterglows
lie on the host galaxies \cite{1998HoggFruchter}. However, our
sample of the observed  GRBs with afterglows is yet limited to
the bursts longer than $6\,$s as BeppoSAX triggers on this
timescale.  Long bursts could be connected with the 
hypernovae-like events and therefore they are closely associated
with the galaxies. Compact object mergers may be connected with
the short bursts, although it has been argued that mass transfer
in the coalescence of compact object may last much longer 
\cite{1998ApJ...503L..53P}.

Our results are consistent with the calculation by
\cite{1998Bloom},  for the case of a massive galaxy  and the
kick velocity distribution width $\sigma_v=200\,$km~s$^{-1}$. We
have verified the dependence of the distribution  of compact
object mergers  on $\sigma_v$. Our results show that for the
case of massive galaxies the escaping fraction weakly depends on
the distribution of kick velocities. We include also  binaries
with objects that are higher mass than the canonical
$1.4\,M_\odot$. These binaries are formed through accretion from
a giant companion onto the neutron star.  In this  calculation
the highest mass of a compact object is below $2.5\,M_\odot$.
The distribution of these more massive binaries is slightly more
concentrated around the galaxies.

It has to be noted that there is a numnber of potential
selection effects which may affect the results of this study.
Assuming that compact objcet mergers are reponsible for GRBs
there may be qualitative differences between the NS-NS mergers
and NS-BH mergers. As indicated above, their spatial 
distribution around host galaxies is different.  Also the
typical timescale of the bursts may be different between these
two classes. Gamma-ray bursts form two separate classes (long
vs. short) with different brightness  distributions and spectra
(short burst are harder than the long ones). So far the study of
afterglows has been possible only for the long bursts. It may be
the case that compact object mergers are connected with the
short bursts for which so far no information about the host
galaxies exist.

The host galaxies have been identified in a long observational
procedure: a gamma-ray burst lead to identification  of  a
fading X-ray source, and then to discovery of the optical
afterglow. Precise observations of the optical afterglows lead
to the disory of host galaxies. There are bursts for which  the
X-ray or optical afterglows were not found. Since afterglows
are  usually connected with external shocks, gamma-ray bursts
that take place outside of galaxies have much weaker afterglows
because of the low density of intergalactic matter. Begelmann
etal. Begelman and Meszaros \shortcite{1993MNRAS.265L..13B}
argue that the afterglow emission depends scales only with the
square root of the density of the outside medium. The mean
external  densities measured from the analysis of the known
afterglow lightcurves  are typically $n\approx 0.03$~cm$^{-3}$
\cite{1998Galama}, while the intergalactic medium may be as
rarified as $10^{-6}$\,cm$^{-3}$. Hence the afterglow of  a
burst taking place outside a galaxy may be up to  two orders of
magnitude weaker than the one in a galaxy.  It shows that there
may be a strong preference against identification of host
galaxies for the bursts that take place outside of galaxies.

Acknowledgments.
 This work has been funded by the following  KBN grants: 2P03D01616
2P03D00911, 2P03D00415 and 2P03D01113, and also made use of the
NASA Astrophysics Data System. TB is grateful for the hospitality of Ecole
Polytechnique where this work was finished.

\newcommand{\apj}{ApJ}
\newcommand{\apjl}{ApJ Lett.}
\newcommand{\apss}{Astr. \& Sp. Sc.}
\newcommand{\bain}{Bull. Astr. Soc. Netherlands}
\newcommand{\iaucirc}{IAU Circ.}
\newcommand{\mnras}{MNRAS}
\newcommand{\pasj}{PASJ}
\newcommand{\aap}{A\&A}
\newcommand{\nat}{Nature}

\label{lastpage}


\begin{thebibliography}{35}
\expandafter\ifx\csname natexlab\endcsname\relax\def\natexlab#1{#1}\fi

\bibitem[\protect\citename{{Begelman} et~al., }1993]{1993MNRAS.265L..13B}
{Begelman} M., {Meszaros} P., and {Rees} M., 1993, \mnras, 265, L13

\bibitem[\protect\citename{Belczynski and Bulik, }1999]{BB1998}
Belczynski K. and Bulik T., 1999, \aap, in press (astro-ph/9901193)

\bibitem[\protect\citename{{Bethe} and {Brown}, }1998]{Bethe1998}
{Bethe} H.A. and {Brown} G.E., 1998, \apj, 506, 780

\bibitem[\protect\citename{{Blaauw}, }1960]{1960BAN....15..265B}
{Blaauw} A., 1960, \bain, 15, 265

\bibitem[\protect\citename{{Blaauw} and {Ramachandran}, }1998]{BlauwRama1998}
{Blaauw} A. and {Ramachandran} R., 1998, astro-ph/9806292

\bibitem[\protect\citename{{Blaes} and {Rajagopal}, }1991]{1991ApJ...381..210B}
{Blaes} O. and {Rajagopal} M., 1991, \apj, 381, 210

\bibitem[\protect\citename{{Bloom} et~al., }1998{\natexlab{a}}]{1998Bloom}
{Bloom} J., {Sigurdsson} S., and O. P., 1998{\natexlab{a}}, astro-ph/9805222

\bibitem[\protect\citename{{Bloom} et~al.,
  }1998{\natexlab{b}}]{1998ApJ...508L..21B}
{Bloom} J.S., {Frail} D.A., {Kulkarni} S.R., et~al., 1998{\natexlab{b}}, \apjl,
  508, L21

\bibitem[\protect\citename{{Bulik} and {Lamb}, }1995]{1995Ap&SS.231..373B}
{Bulik} T. and {Lamb} D.Q., 1995, \apss, 231, 373

\bibitem[\protect\citename{{Bulik} et~al., }1998]{1998ApJ...505..666B}
{Bulik} T., {Lamb} D.Q., and {Coppi} P.S., 1998, \apj, 505, 666

\bibitem[\protect\citename{{Cordes} and {Chernoff}, }1997]{1997ApJ...482..971C}
{Cordes} J.M. and {Chernoff} D.F., 1997, \apj, 482, 971

\bibitem[\protect\citename{{Costa} et~al., }1997]{1997IAUC.6576....1C}
{Costa} E., {Feroci} M., {Piro} L., et~al., 1997, \iaucirc, 6576, 1

\bibitem[\protect\citename{{Djorgovski}, }1999]{GCN256}
{Djorgovski} S.G. e., 1999, GCN, 256

\bibitem[\protect\citename{{Fruchter}, }1998]{1998Fruchter}
{Fruchter} A., 1998, Talk at the Texas Symposium, Paris 1998

\bibitem[\protect\citename{{Galama} and {Wijers}, }1998]{1998Galama}
{Galama} T. and {Wijers} R., 1998, astro-ph/9805341

\bibitem[\protect\citename{{Groot} et~al.,
  }1997{\natexlab{a}}]{1997IAUC.6584....1G}
{Groot} P.J., {Galama} T.J., {van Paradijs} J., et~al., 1997{\natexlab{a}},
  \iaucirc, 6584, 1

\bibitem[\protect\citename{{Groot} et~al.,
  }1997{\natexlab{b}}]{1997IAUC.6588....1G}
{Groot} P.J., {Galama} T.J., {Van Paradijs} J., et~al., 1997{\natexlab{b}},
  \iaucirc, 6588, 1

\bibitem[\protect\citename{{Hansen} and {Phinney}, }1997]{1997MNRAS.291..569H}
{Hansen} B.M.S. and {Phinney} E.S., 1997, \mnras, 291, 569

\bibitem[\protect\citename{{Hogg} and {Fruchter}, }1998]{1998HoggFruchter}
{Hogg} D. and {Fruchter} A., 1998, astro-ph/9807262

\bibitem[\protect\citename{{Iben} and {Tutukov}, }1996]{1996ApJ...456..738I}
{Iben} Icko J. and {Tutukov} A.V., 1996, \apj, 456, 738

\bibitem[\protect\citename{{Klebesadel} et~al., }1968]{Klebesadel68}
{Klebesadel} R., {Strong} X., and Y. O., 1968, \apj, 10, 000

\bibitem[\protect\citename{{Klu{\'z}niak} and {Lee},
  }1998]{1998ApJ...494L..53K}
{Klu{\'z}niak} W. and {Lee} W.H., 1998, \apjl, 494, L53

\bibitem[\protect\citename{{Klu{\'z}niak} and {Ruderman},
  }1998]{1998ApJ...505L.113K}
{Klu{\'z}niak} W. and {Ruderman} M., 1998, \apjl, 505, L113

\bibitem[\protect\citename{{Kuijken} and {Gilmore}, }1989]{1989MNRAS.239..571K}
{Kuijken} K. and {Gilmore} G., 1989, \mnras, 239, 571

\bibitem[\protect\citename{{Kulkarni} et~al., }1998]{1998Natur.393...35K}
{Kulkarni} S.R., {Djorgoski} S.G., {Ramaprakash} A.N., et~al., 1998, \nat, 393,
  35

\bibitem[\protect\citename{{Miyamoto} and {Nagai}, }1975]{1975PASJ...27..533M}
{Miyamoto} M. and {Nagai} R., 1975, \pasj, 27, 533

\bibitem[\protect\citename{{Narayan} et~al., }1992]{1992ApJ...395L..83N}
{Narayan} R., {Paczynski} B., and {Piran} T., 1992, \apjl, 395, L83

\bibitem[\protect\citename{{Paczy{\'n}ski}, }1990]{1990ApJ...348..485P}
{Paczy{\'n}ski} B., 1990, \apj, 348, 485

\bibitem[\protect\citename{{Paczynski}, }1998]{1998ApJ...494L..45P}
{Paczynski} B., 1998, \apjl, 494, L45

\bibitem[\protect\citename{{Peters}, }1964]{Peters1964}
{Peters} P.C., 1964, Phys.~Rev., 136, 1224

\bibitem[\protect\citename{{Pols} and {Marinus}, }1994]{1994A&A...288..475P}
{Pols} O.R. and {Marinus} M., 1994, \aap, 288, 475

\bibitem[\protect\citename{{Portegies Zwart}, }1998]{1998ApJ...503L..53P}
{Portegies Zwart} S.F., 1998, \apjl, 503, L53

\bibitem[\protect\citename{{Sahu} et~al., }1997]{1997ApJ...489L.127S}
{Sahu} K.C., {Livio} M., {Petro} L., et~al., 1997, \apjl, 489, L127

\bibitem[\protect\citename{{Tutukov} and {Yungelson},
  }1993]{1993MNRAS.260..675T}
{Tutukov} A.V. and {Yungelson} L.R., 1993, \mnras, 260, 675

\bibitem[\protect\citename{{Vrancken} et~al., }1991]{1991A&A...249..411V}
{Vrancken} M., {De Greve} J.P., {Yungelson} L., and {Tutukov} A., 1991, \aap,
  249, 411

\end{thebibliography}
\end{document}